\documentclass[fleqn,usenatbib]{mnras}

\usepackage{newtxtext,newtxmath}

\usepackage[T1]{fontenc}
\usepackage{ae,aecompl}

\usepackage{graphicx}      
\usepackage{amsmath}       
\usepackage{amssymb}       
\usepackage{braket}        
\usepackage{enumerate}     
\usepackage{dsfont}        

\newcommand{\plh}{{\ooalign{$\phantom{0}$\cr\hidewidth$\scriptstyle\times$\cr}}}
\DeclareRobustCommand{\imdim}[1]{$#1 \plh #1$}
\DeclareMathOperator*{\argmin}{\arg\!\min}

\newcommand\blfootnote[1]{%
  \begingroup
  \renewcommand\thefootnote{}\footnote{#1}%
  \addtocounter{footnote}{-1}%
  \endgroup
}

\makeatletter
\def\@printed{
    \qquad\qquad\qquad Compiled using MNRAS \LaTeX\ style file v\@version}
\gdef\@journal{\hfill\@printed}
\def\bsp{}
\def\@oddfoot{}
 \def\@evenhead{}
 \def\@evenfoot{}
\def\ps@titlepage{\let\@mkboth\@gobbletwo
 \def\@oddhead{\footnotesize\@journal}
 \def\@oddfoot{\hfil}
 \def\@evenhead{\hfill}
 \def\@evenfoot{\hfil}
 \def\sectionmark##1{}
 \def\subsectionmark##1{}}
\def\abstract{\if@twocolumn
   \start@SFBbox\@abstract
 \else
   \@abstract
 \fi}
\def\endabstract{\if@twocolumn
   \endlist\finish@SFBbox
 \else
  \endlist
 \fi}
\def\@abstract{\list{}{%
    \listparindent\realparindent
    \itemindent\z@
    \labelwidth\z@ \labelsep\z@
    \leftmargin 1.5pc\rightmargin 1.5pc
    \parsep 0pt plus 2pt}\item[]%
    \reset@font\normalsize{\bf ABSTRACT}\\\reset@font\large
}
\def\@maketitle{\newpage
 \vspace*{7pt}
 {\raggedright \sloppy
  {\reset@font\huge \bf \@title \par}
  \vskip 23pt
  {\reset@font\LARGE
   \begin{tabular}[t]{@{}l@{}}\let\\=\author@nextline\@author
   \end{tabular}
   \par}
  \vskip 22pt
 }
 \par\noindent
 {\reset@font\small \par}
 \vskip 22pt
}
\makeatother

\title[High-resolution synthetic galaxies with GANs]{Forging new worlds: high-resolution synthetic galaxies with chained generative adversarial networks}

\author[Fussell, L. and Moews, B.]{
Levi Fussell,$^{1}$\thanks{E-mail: \href{mailto:levi.fussell@ed.ac.uk}{levi.fussell@ed.ac.uk}}
Ben Moews,$^{2}$
\\
$^{1}$Institute of Perception, Action and Behaviour, University of Edinburgh, 10 Crichton St, Edinburgh EH8 9AB, UK\\
$^{2}$Institute for Astronomy, University of Edinburgh, Royal Observatory, Edinburgh EH9 3HJ, UK\\
}

\date{Accepted 2019 February 25. Received 2019 February 22; in original form 2018 December 6}

\pubyear{2019}

\begin{document}
\label{firstpage}
\pagerange{\pageref{firstpage}--\pageref{lastpage}}
\maketitle

\begin{abstract}
Astronomy of the 21st century increasingly finds itself with extreme quantities of data. This growth in data is ripe for modern technologies such as deep image processing, which has the potential to allow astronomers to automatically identify, classify, segment and deblend various astronomical objects. In this paper, we explore the use of chained generative adversarial networks (GANs), a class of generative models that learn mappings from latent spaces to data distributions by modelling the joint distribution of the data, to produce physically realistic galaxy images as one use case of such models. In cosmology, such datasets can aid in the calibration of shape measurements for weak lensing by augmenting data with synthetic images. By measuring the distributions of multiple physical properties, we show that images generated with our approach closely follow the distributions of real galaxies, further establishing state-of-the-art GAN architectures as a valuable tool for modern-day astronomy.
\end{abstract}


\raggedbottom

\section{Introduction}
\label{sec:intro}

\blfootnote{
\hrule{}
\vspace{5pt}
\hspace{-10pt}
This is a pre-copyedited, author-produced PDF of an article accepted for publication in \textit{Monthly Notices of the Royal Astronomical Society} following peer review. The version of record, Fussell, L. and Moews, B. (2019), ``Forging new worlds: high-resolution synthetic galaxies with chained generative adversarial networks'', \textit{MNRAS}, 485(3):3203-3214, is available online at: \url{https://doi.org/10.1093/mnras/stz602}.}Interest in using machine learning for tasks such as galaxy processing, classification, segmentation, and deblending has become popular due to the growth of larger galaxy datasets \citep{galaxyClass3, galaxyClass1, galaxyClass2, deblendingGAN}. As these approaches become more complex and attempt to automate galaxy pre-processing, the data requirements grow accordingly. Galaxy datasets such as Galaxy Zoo data releases, as described by \citet{galaxyZoo} and \citet{galaxyZoo2}, and the EFIGI catalogue by \citet{efigiData} are examples of large, pre-processed datasets used to train machine learning models \citep{gzMl2, gzMl1}.

Currently, much research is invested in how to use generative adversarial networks (GANs), a type of unsupervised machine learning model featuring two neural networks that compete in a zero-sum game, for purposes other than subjectively beautiful visuals. Strong generative models that have generalised well to a training dataset will have learned useful features about the true data distribution. In the context of galaxies, this can involve physical properties such as ellipticity, brightness, size, shape, and colour. Galaxy datasets can then be indefinitely enlarged by the generative model. Deep object segmentation and image classification, which require a large number of training examples, are examples of applications that benefit from the training with such synthetic datasets. These improved models can be used to more accurately segment and identify galaxies from telescope data. 

Datasets augmented in this way also solve the need for large quantities of high-quality datasets of galaxy images in 21st-century cosmology. One example is in the field of weak lensing: when trying to differentiate between different dark energy models using galaxy distortions, reducing biases in galaxy shape measurements is of importance. To correctly extract the lensing signal, simulations are required to calibrate shape measurement algorithms. High-quality real galaxy images used for this calibration are, however, costly to obtain. This problem is especially relevant to upcoming cosmological probes such as LSST and Euclid \citep{lsst, euclid}.

While research on the generative modelling of galaxies with machine learning is sparse in the related literature, recent efforts explore this application case. \citet{darkGan} train both a conditional variational autoencoder (C-VAE) and a conditional generative adversarial network (C-GAN) to create galaxy images with a resolution of \imdim{128} pixels based on conditioning features from the training data, for example brightness and size of galaxies. C-VAEs are a type of neural network that compress the $N$-dimensional representations of a dataset into smaller $L$-dimensional latent representations, where the latent variables are a multivariate Gaussian distribution. C-GANs, on the other hand, are generative adversarial networks that generatively model a target dataset based on a set of conditional inputs in addition to the usual latent noise input. The general GAN model is discussed further in Section~\ref{sec:generativeadversarialnetworks}. The aim of their analysis is to provide datasets for shape measurement algorithms in the cosmological probes mentioned above. The authors report that by conditioning their models on the features of a galaxy from the real dataset, they are able to successfully reproduce rich galaxy images that share similar structures with real data. In order to quantify the similarity of generated images to real data, the ellipticity and size distributions of the real and generated galaxies are measured. The attempt to train a C-GAN on continuous conditional variables, however, is reported to fail, and \citeauthor{darkGan} replace the discriminator network with a novel predictor network producing desirable convergence properties for the generator. In related research, but targeting another type of image relevant to cosmology, \citet{cosmoWeb} use GAN models to generate synthetic cosmic web examples, with similar statistical evaluations of the results.

\citet{ganRecovery} make use of a different capability of GANs to recover high-quality image features from artificially degraded galaxies. The applications for such an approach to astronomy are clear, as galaxy images from telescopes suffer from various noise sources such as background noise, atmospheric noise, and instrumental noise, which convolve and degrade the image. Generative models that are able to automatically filter this noise and reproduce rich galaxy images can help to streamline the galaxy-imaging process. The authors use a relatively small dataset of 4,105 galaxy images for training, and thus encounter difficulties with anomalous galaxies, for example a tidally warped edge-on disc galaxy, due to a lack of generalisation in their model.

In a third take on using GANs, \citet{deblendingGAN} propose a branched GAN to deblend overlapping galaxies in compound images in the $g$, $r$, and $i$ bands, with each branch generating one of the two separated galaxies. Using a model based on the super-resolution GAN (SRGAN) previously introduced by \citet{superRes1}, the authors exploit the ability of GANs to fill in missing pixels occluded by the superposition of blended galaxies. One of the advantages of GANs and similarly pretrained architectures is the speed of the fully automated task, which offers a way to avoid the discarding of blended galaxy images due to severe blends.

The rest of this paper is structured as follows: Section~\ref{sec:generativeadversarialnetworks} provides a brief outline of the general GAN model and an introduction to the variations used in our experiment, the deep convolutional GAN (DCGAN) and the stacked GAN (StackGAN). Descriptions of the model architectures, training schedule, dataset, and hardware are included for reproducibility. Next, Section~\ref{sec:experiments} outlines the experiments performed with the DCGAN and StackGAN architectures to qualitatively evaluate design decisions of the model architectures. In Section~\ref{sec:eval}, we perform a closest-match analysis to rule out simple training set memorisation by our model, and quantitatively evaluate the generated galaxies of the best-performing models for a DCGAN producing \imdim{64} images and a chained DCGAN/StackGAN producing higher-resolution \imdim{128} images. This evaluation involves measuring the ellipticity, angle from the horizon, total flux, and size as measured by the semi-major axis, for both real and generated galaxy images, thus showing that the generated image distributions closely follow the real image distributions. Finally, we provide a discussion of the results and remarks on future work in Section~\ref{sec:discussion}, and final conclusions in Section~\ref{sec:conclusion}.

\section{Generative adversarial networks}
\label{sec:generativeadversarialnetworks}

We provide a brief outline of the basic GAN architecture, colloquially referred to as \textit{vanilla} GANs from here on, in Section~\ref{gandesign}. For a more extensive introduction to GANs, we refer the interested reader to \citet{gan} and \citet{principleGan}. We extend this introduction by covering the specifics of deep convolutional GANs in Section~\ref{sec:dcgan}, and introduce the StackGan architecture for later experiments in Section~\ref{stackgan}.

\subsection{Basics of GAN design}
\label{gandesign}

The vanilla GAN architecture consists of two neural networks, the \textit{generator} and the \textit{discriminator}, which have adversarial objectives. The generator's objective is to `trick' the discriminator by generating fake data that is close to the real data distribution, while the discriminator's objective is to determine if the data it is presented with is drawn from the real or fake data distribution. This can be represented by the following two-player minimax game:
\begin{eqnarray}
\begin{aligned}
\label{eq:gan}
\min\limits_G\{ \max\limits_D \{ &\mathbb{E}_{x \sim X_{\mathrm{real}}}[\log D(x)] \\
 & + \mathbb{E}_{z \sim \mathbb{N}(0, \mathds{I}_{\sigma})}[\log(1 - D(G(z)))] \} \}
\end{aligned}
\label{eqn}
\end{eqnarray}
Here, $D(\cdot)$ is the discriminator function, which takes as its input a data sample $x \sim X_{\mathrm{real}}$ or $x \sim X_{\mathrm{fake}}$ and outputs the probability $p(x \in X_{\mathrm{real}} \ | \ x)$ of the data sample belonging to the real data distribution. $G(\cdot)$ is the generator function, which takes as its input randomly sampled multivariate Gaussian noise $z \sim \mathbb{N}(0, \mathds{I}_{\sigma})$, with $\sigma \in \mathbb{R}_{>0}$ and $\mathds{I}_{\sigma} \in \mathbb{R}^{100} \times \mathbb{R}^{100}$ as the identity matrix with diagonal values $\sigma$, and outputs a data sample $x \sim X_{\mathrm{fake}}$ that is as close to the true data distribution as possible. During training, it is common to maximise the alternative objective function $\log(D(G(z)))$ for the generator, as this approach leads to a more stable convergence, while the original objective function is still maximised for the discriminator.

Training GANs requires alternating between training the generator and training the discriminator. This allows for each network to incrementally improve such that both networks seek an optimal equilibrium. Although convergence should occur in theory, in practise, GANs often struggle from imbalanced player strengths, mode collapse, and oscillations, which we explore in Section~\ref{sec:dcganexperiments}.

\subsection{Deep convolutional GAN}
\label{sec:dcgan}

\begin{figure*}
\includegraphics[width=\textwidth]{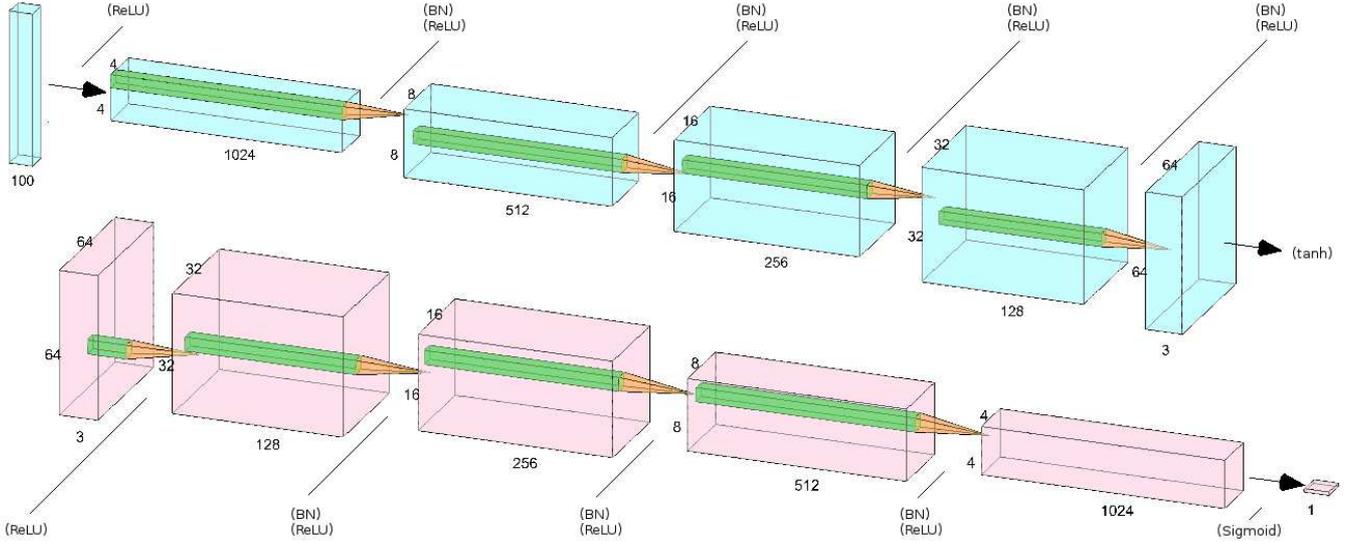}
\centering
\caption{DCGAN architecture employed in this paper. The top (blue) CNN is the generator that is tasked with creating realistic synthetic galaxy images from Gaussian noise, while the bottom (red) CNN is the discriminator, the purpose of which is to learn to differentiate between generated and real images, forcing the generator to learn how to produce more realistic galaxy images. Both networks employ batch normalisation (BN) and the rectified linear unit function (ReLU), with the generator and discriminator using the hyperbolic tangent function (tanh) and the sigmoid function (Sigmoid), respectively. See Appendix~\ref{app:appA} for function definitions and Section~\ref{sec:batchnorm} for batch normalisation.}
\label{fig:dcgan}
\end{figure*}

Because vanilla GANs work well for image generation, a natural extension is to use convolution layers, as first introduced by \citet{cnn}, which decrease the number of parameters per layer via weight sharing of convolving templates:
\begin{eqnarray}
\mathrm{output} = \varphi \left( b + \sum^K_{l = 0} \sum^K_{m = 0} \sum^C_{n = 0} w_{l, m, n} a_{i + l, j + m, k + n} \right)
\end{eqnarray}
Here, the output for a hidden neuron at location $(i, j, k)$ is described by an activation function $\varphi$, a shared bias $b$, an input activation value $a$, and $K^2 C$ as the number of shared weights $w$ for the local receptive field of size \imdim{K} with $C$ channels. Neurons, in this context, are elementary input-output functions with an \textit{activation function}, which receive one or multiple inputs to provide summed-up results of that activation function. Conveniently, these reduced-parameter templates are also location-invariant when convolved across the input, which is an essential requirement for effective object recognition. Each layer $l$ is represented by a $C_{l}\times K\times K \times C_{l+1}$ tensor, where $C_{l}$ is the input channel size, or \textit{channels}, $C_{l+1}$ is the output channel size and the number of convolutions, and $K$ is the area of the $C_{l} \times K \times K$ convolution template, which is also referred to as a \textit{kernel}. \citet{dcgan} present the first working DCGAN. The fundamental insights in their work include the removal of all pooling layers, which are  used in traditional convolutional neural networks (CNNs), and their replacement with more convolution layers, as well as the addition of batch normalisation layers, the use of the ReLU activation function introduced by \citet{relu} for the generator, and the use of the LeakyReLU activation function for the discriminator \citep{leakyReLU}. 
The DCGAN architecture used in this paper is similar to the original architecture outlined in \citet{dcgan}. We provide a schematic representation in Figure~\ref{fig:dcgan}, and further details on the architecture in Appendix~\ref{app:appA}.

\subsection{StackGAN}
\label{stackgan}

The StackGAN architecture is introduced by \citet{stackGan}, but has since been developed in further versions, for example an attention-based model by \citet{stackGanAtt} and StackGAN++ by \citet{stackGan++}. Due to basic GAN architectures not scaling well to image sizes larger that \imdim{64}, the StackGAN architecture employs two GANS; one to generate low-resolution \imdim{64} synthetic images as the DCGAN does, and another one to transform the synthetic images into high-resolution versions. Research on generating high-quality realistic images using GANs has recently become popular with a variety of proposed architectures \citep{superRes1, superRes2, superRes3}. This subdomain is called \textit{super resolution} and differs from the StackGAN model in that realism to the human eye is targeted, without necessarily requiring data that is similar to the true data distribution. 

In this paper, we use an architecture similar to StackGAN, which is defined by two GANs, the Stage-I GAN and the Stage-II GAN. The Stage-I GAN generates low-resolution images, while the Stage-II GAN converts them into higher-resolution images. Both models are trained independently. The original StackGAN architecture uses nearest-neighbours upsampling layers coupled with \imdim{3} kernel convolutions in the Stage-I generator, and conditions the GAN on an embedded text describing the input image. In this work, the Stage-I generator is replaced by the DCGAN generator architecture as described in Section~\ref{sec:dcgan}, which is used to generate lower-resolution images. For the Stage-II GAN, we use an architecture similar to the StackGAN by \citet{stackGan}, but also incorporate elements from the architecture in \citet{superRes1} inspired by StackGAN. The novelties of the Stage-II generator are downsampling layers for feature extraction, residual connections to preserve low-level information of pixels, and nearest-neighbours upsampling with 3x3 convolutions to encourage the resolution growth of the image as it passes through the generator.

To encourage the generator to produce an image similar to the real high-resolution target image, we also introduce a pixel-loss term into the generator objective function. We refer to this new generator objective function as the \textit{dual-objective function}, which is used in a super-resolution GAN model by \citet{superRes1}. The latter authors define one of the terms as the `content loss', which computes an error metric between the resolution-enhanced image and the real high-resolution image, and the second term as the `generative loss', which is the loss based on the discriminator's output. Without the use of the dual-objective function, we find that during experiments the generator focusses on producing galaxy images related to the galaxy types most common in the distribution. By enforcing pixel-to-pixel similarity on the upscaled image, the generator produces high-resolution images retaining rarer characteristics of the galaxies, for example spiral arms. The dual-objective function we use for the Stage-II generator is, therefore, given by:
\begin{eqnarray}
\label{eq:dual_obj}
L_{G}(x_{\mathrm{real}}, x_{\mathrm{fake}}) = \lambda_D L_{D}(x_{\mathrm{fake}}) + \lambda_P L_{P}(x_{\mathrm{real}}, x_{\mathrm{fake}})
\end{eqnarray}
$L_{D}=-\log(D(G(z)))$ is the usual generative loss term based on the discriminator output given the generated fake image $x_{\mathrm{fake}}$ (see equation~\ref{eq:gan}), $L_{P}$ is the pixel-loss term given the real image $x_{\mathrm{real}}$ and its generated resolution-enhanced counterpart $x_{\mathrm{fake}}$, and $\lambda_D, \lambda_P \in \mathbb{R}_{> 0}$. The pixel-loss term is defined as:
\begin{eqnarray}
L_{P}(x_{\mathrm{real}}, x_{\mathrm{fake}}) = \frac{1}{D^2} \sum_{i=1}^D \sum_{j=1}^D \left( x_{\mathrm{real}}^{(i,j)} - x_{\mathrm{fake}}^{(i,j)} \right)^2
\end{eqnarray}
Here, $x^{(i,j)}=R^{(i,j)}+G^{(i,j)}+B^{(i,j)}$ is the sum of the red, green, and blue channel values at pixel position $(i, j)$ for an image $x \in \mathbb{R}^D \times \mathbb{R}^D$. Following \citet{superRes1}, we set $\lambda_D = 0.001$ and $\lambda_P = 1$. Additionally, to ensure that the generator learns latent features of the input images, the first layer of the generator reduces the image size from \imdim{64} to \imdim{32} to enforce an information bottleneck. The architecture for the Stage-II generator is outlined in further detail in Appendix~\ref{app:appB}.

The architecture for the discriminator is identical to the DCGAN discriminator described in Section~\ref{sec:dcgan}, with the exception of a training batch size of $B = 64$ for both the generator and the discriminator, and with an additional layer in the discriminator to transform \imdim{128} images instead of \imdim{64} images into a single output.

\vspace{10pt}

\section{Experiments}
\label{sec:experiments}

In Section~\ref{sec:setupanddataset}, we introduce the setup of our models and the construction of the dataset, followed by descriptions of a variety of DCGAN experiments in Section~\ref{sec:dcganexperiments}. These experiments cover alterations to architecture parameters, which include the kernel size and the number of convolution channels, as well as the use of batch normalisation, label smoothing, and dropout. We also present a closest-match analysis to ensure that the model does not memorise and reproduce the training data. Qualitatively, the architecture produces suitable outputs even for slight variations in most of its parameters, the exception being that batch normalisation layers are essential for the desired level of performance. The generation of higher-resolution \imdim{128} images is presented in Section~\ref{sec:generating128}, followed by the description of a chained approach using StackGANs in Section~\ref{sec:stackganexperiments}.

\subsection{Setup and dataset}
\label{sec:setupanddataset}

The models are trained on a dataset $X = \{x_i \in \mathbb{R}^{D} \times \mathbb{R}^{D} \ | \ 0 < i < N\}$, where $N = 61,578$ is the size of the dataset and $D \times D$ is the resolution of an image. The images are full-colour RGB galaxy images from the Galaxy Zoo 2 data release \citep{galaxyZoo2}. This dataset is a set of images that have been centred, cropped, and resized to resolution $D=424$ such that a single galaxy is found at its centre. The presence of foreground stars, background noise, and extraneous galaxies in the images provides a further test for the ability of our approach. Unlike \citet{darkGan}, we do not crop the images by $50\%$ to reduce the effects of this noise, thus allowing our work to be used as a more challenging measure of the capabilities of GANs in this context.

The models are trained on a single NVIDIA 1060 GTX 6GB GPU. During training, the galaxy images are flipped horizontally and vertically with probability $0.5$. A fixed mini-batch size $B \in \{32, 64, 128\}$ is selected for the entire training process and each mini-batch is randomly sampled from the dataset, and an entire epoch is complete when all batches of the dataset have been sampled. Both networks are trained in an alternating manner with the Adam optimiser by \citet{adam}, where one network is trained for one step of gradient descent before switching to train the other network. The learning parameters and weight initialisation method are outlined in Table~\ref{tab:params}. 

\begin{table}
\centering
\caption{Learning parameters and weight initialisation method for DCGAN and StackGAN training. For the StackGAN, the learning rate is scheduled to halve every $80$ epochs, where $e$ represents the current epoch.}
\label{tab:params}
\begin{tabular}{lcc}
\hline
& DCGAN & StackGAN \\ \hline
learning rate &	$0.0002$ & $0.0002 \times 0.5^{\lfloor \frac{e}{80}\rfloor}$ \\
$\beta_1$ &	$0.5$ &	$0.5$ \\ 
$\beta_2$ &	$0.999$	&	$0.999$	\\ 
weight initialisation &	$\mathbb{N}(0,0.02)$ &	$\mathbb{N}(0,0.02)$ \\	\hline
\end{tabular}
\end{table}

Since the DCGAN performance is known to drop for images larger than \imdim{64}, as shown by \citet{improveGan}, initial experiments focus on generating \imdim{64} galaxy images, whereas results for generating \imdim{128} images are described later. The \imdim{64} images for real data are created by downscaling the original \imdim{424} images using nearest-neighbour downsampling. In a subsequent step, we enhance the generated images to a resolution of \imdim{128} using a StackGAN.

\subsection{DCGAN experiments}
\label{sec:dcganexperiments}

We first perform architectural experiments to explore how adjustments to the DCGAN affect the generative performance. These experiments include changing the kernel size of the convolutions, changing the number of convolutional channels, removing batch normalisation layers, adding label smoothing to the objective function, and including dropout layers. The separate experiments are described in further detail below, and the results are presented. Each experiment starts with the previously described DCGAN architecture and adjusts the parameter that is the focus of the experiment. Evaluation of the results during these experiments is qualitative, and the best results during this stage are chosen for quantitative evaluation in Section~\ref{sec:propertydistros}.

\subsubsection{Kernel size}
\label{sec:kernelsize}
\begin{figure}
\includegraphics[width=\columnwidth]{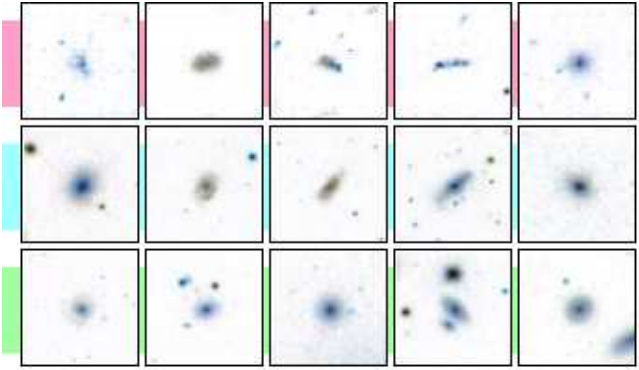}
\centering
\caption{Comparison of the effect of different kernel sizes on the resulting images. The first row (red) shows generated images for a kernel size of $K=6$, the second row (blue) shows generated images for a kernel size of $K=4$, and the third row (green) shows images from the Galaxy Zoo dataset. Images are randomly sampled.}
\label{fig:kerExp}
\end{figure}

Each channel of a convolutional layer represents a $K\times K \times C$ kernel with $K, C \in \mathbb{N}$, where $K$ is the size of the kernel and $C$ is the channel size as described in Section~\ref{sec:dcgan}. The kernel  performs a weighted sum of the pixels within the kernel region as it convolves over the image. More formally, if $k(x, y, z)$ represents a kernel function applied to a pixel $q^{(x,y,z)}$ at location $(x,y,z)$ in the image, the convolution can be defined as:
\begin{eqnarray}
q * k = \sum_{i = 0}^{K - 1}\sum_{j = 0}^{K - 1}\sum_{m = 0}^{C - 1} k(i, j, m)q^{(x + i, y + j, z + m)}
\end{eqnarray}
The kernel moves along the image at a per-pixel rate $s \in \mathbb{N}_{> 0}$ called the \textit{stride}. A kernel of size $k$ will have an amount of overlap $k-s$; when this value is positive, the kernel regions overlap by $|k-s|$; when this value is negative, there are $|k-s|$ gaps between kernel regions. A constant stride value of $s=2$ is used while exploring the use of a larger kernel of size $K=6$, which means that there is an overlap of $4$ between kernel regions. The results are compared to a $K=4$ kernel model with the same stride and an overlap of $2$. Both models are run for $100$ epochs, and the results are shown in Figure~\ref{fig:kerExp}. The $K=6$ kernel model produces asymmetrical galaxies and surrounding objects, as well as sharp cut-off boundaries due to the overlapping of the kernel as it convolves the image.

\subsubsection{More convolution channels}

Models with more convolutional channels, as described in Section~\ref{sec:kernelsize}, are able to learn a richer set of templates for producing images. In the following experiments, we explore channel scales of $W \in \{32, 64, 128\}$. The DCGAN architecture progressively halves the number of channels in each generator layer, and doubles the number in each discriminator layer. Therefore, the channel sizes we explore are $C \in \{W \times 2^l \ | \ W \in \{32, 64, 128\}\}$, where $l \in \mathbb{N}$ with $0 \leq l < L$ is the index of the layer, and $L$ is the number of layers in the network. Each model is trained for $100$ epochs with a batch size of $B=128$. In addition, a $W=128$ model with a batch size of $B=32$ is explored.

The model with $W=128$ and $B=32$ generates spiral arms in galaxies, whereas previous models struggle with generating such structures. In general, increasing the number of channels improves the quality of the generated galaxies, as the representational power of models with more channels is larger.

\subsubsection{Batch normalisation}
\label{sec:batchnorm}

Batch normalisation, introduced by \citet{batchNorm}, has shown to improve the generalisation of the generator and prevent mode collapse \citep{dcgan}. A model that has good generalisation is capable of performing well on unseen data. Mode collapse is an unsolved problem when training GANs and describes the event in which the generator `collapses', outputting only the mode of the distribution it is trying to model. Once collapsed, the generator cannot usually recover because the gradients near a collapsed distribution approach zero. Also, adding batch normalisation to the discriminator helps with gradient computation by keeping the gradients between layers independent. Batch normalisation is placed after the layer weights, but before the layer activation, in order to normalise the input batch. It normalises the minibatch as it passes through a layer according to statistics computed from the minibatch:
\begin{eqnarray}
x'_{i,j} = \frac{x_{i,j} - \braket{x_j}}{\sqrt{\braket{x_j - \braket{x_j}}^2 + \epsilon}}
\end{eqnarray}
Here, $\braket{x_j}$ represents the mean of the $j^{\mathrm{th}}$ component of $x_i \in X$, $X=\{x_i\in \mathbb{R}^N \ | \ 0 < i < B\}$ denotes the mini-batch of size $B$, and $\epsilon \ll 1$. The layer then learns parameters $\alpha$ and $\beta$ via gradient descent to control the normalisation factor for the entire mini-batch via $BN(x'_i)=\alpha x_i + \beta$.

We exclude the discriminator output and generator input from the addition of batch normalisation layers to avoid oscillations \citep{dcgan}. Our results support the claim that batch normalisation improves GAN performance. In contrast, the model without batch normalisation between the layers lacks both colour and diversity in the range of generated images.

\subsubsection{Label smoothing}

Instead of writing the optimisation criteria under the entire data distribution, Equation~\ref{eq:gan} can be used to write the loss of a single data point $x$ under the discriminator as:
\begin{eqnarray}
\begin{aligned}
\label{eq:gansingle}
t_{x}(\log(D(x))) + (1-t_{x})(1-\log(D(x))
\end{aligned}
\end{eqnarray}
Here, $x$ is a data point from either the real $X_{\mathrm{real}}$ or fake $X_{\mathrm{fake}}$ data sets. The \textit{label} $t_x$ for $x$ is $t_x=1$ if $x \in X_{\mathrm{real}}$, and $t_x = 0$ if $x \in X_{\mathrm{fake}}$. 
Label smoothing perturbs the labels of the real images fed into the discriminator by $\pm\delta$, with a common choice of $\delta=0.3$. Instead of setting the label of real images to one, a label is sampled uniformly from the range $[1-\delta,1+\delta]$. Previous research shows that the addition of label smoothing leads to a noticeable gain in GAN performance \citep{improveGan}.

Label smoothing appears to remove clutter in the image and reduce the effect of noise on the generator. It also seems to decrease the diversity of colour and shape, tending more of the galaxies towards less elliptical shapes. The advantage of label smoothing is, therefore, not clear.

\subsubsection{Dropout}
\label{sec:dropout}

\begin{figure}
\includegraphics[width=\columnwidth]{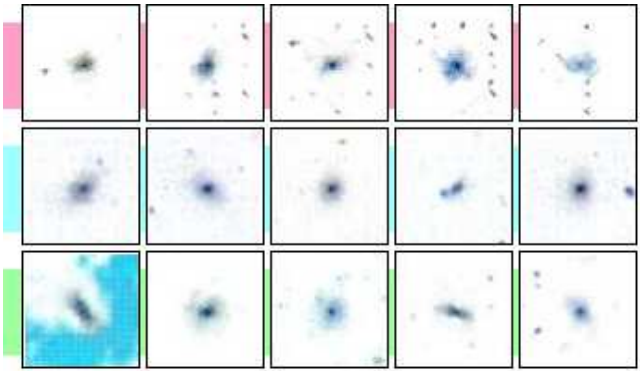}
\centering
\caption{Comparison of the effect of dropout on the resulting images. The first row (red) shows generated images with dropout layers in the generator, the second row (blue) shows generated images with dropout layers in the discriminator, and the third row (green) shows generated images with dropout layers in both the generator and discriminator. Images are randomly sampled.}
\label{fig:dropoutExp}
\end{figure}

A dropout layer is represented by a Bernoulli distribution $\mathrm{Bernoulli}(\varphi)$ for a neuron $\varphi$ in the network. With probability $p$, a neuron in the previous layer will output $\mathrm{out}(\varphi)=0$ during training; otherwise, its output remains unchanged \citep{dropout}. During testing, the output of a neuron is then an expectation under the Bernoulli distribution such that:
\begin{eqnarray}
\mathbb{E}_{p}[\mathrm{out}(\varphi)] = p \cdot \mathrm{out}(\varphi)
\end{eqnarray}
A value of $p=0.5$ is used for all experiments, and a variation of dropout called \textit{spatial dropout} is employed, as our networks are fully convolutional \citep{dropoutSpatial}. Dropout helps the generator or discriminator to generalise better and not overfit the data. The placement of dropout layers is explored by adding dropout layers between all hidden layers of the generator, all hidden layers of the discriminator, and all hidden layers of both the generator and discriminator. The results are shown in Figure~\ref{fig:dropoutExp}.

Adding dropout layers in the generator causes poor performance, leading to visually unrealistic images. With dropout in the discriminator, the generator is still able to produce realistic galaxies, but the discriminator's strength is decreased, meaning that it is easier for the generator to overcome the discriminator, thus making generated images less diverse. Dropout in both the generator and discriminator still results in poor performance for the generator. In general, dropout does not appear to benefit the model. Despite this result, we find that a single, carefully-managed dropout layer on the discriminator can help for larger image generation, which is further discussed in Section~\ref{sec:generating128}.

\subsection{Generating 128x128 images}
\label{sec:generating128}

The DCGAN architecture was not originally designed to handle \imdim{128} images and, as mentioned previously, images of \imdim{128} and above pose a challenge for simple GAN models. Therefore, scaling up the \imdim{64} image architecture to work with a resolution of \imdim{128} is not a straightforward task, and we observe that adding a sixth convolutional layer results in mode collapse due to the discriminator being too powerful. During these tests, we use two techniques to try to weaken the discriminator: placing a dropout layer before the final discriminator layer, and decreasing the number of channels in the discriminator.

The best model requires some unconventional training methods. In Appendix~\ref{app:appC}, we list the details of the the final DCGAN model used for generating \imdim{128} images, which differs slightly from the DCGAN architecture described in Section~\ref{sec:dcgan}.

Training with a dropout layer diminishes the discriminator's performance early on, eventually causing mode collapse. Alternatively, training without dropout creates a discriminator that proves too powerful after $200$ epochs of training. By introducing a dropout layer at the $200^{\mathrm {th}}$ epoch, however, the discriminator is provided with enough time to learn a good function at first, but is then weakened to allow the generator to develop more generative power. Other architectures we experiment with either mode-collapse, produce images that are too bright, or show obvious kernel templates throughout the model image. For the best model, the generator produces  desirable galaxy images after $400$ epochs of training, but the background still shows evident kernel template artifacts. 

\begin{figure}
\includegraphics[width=\columnwidth]{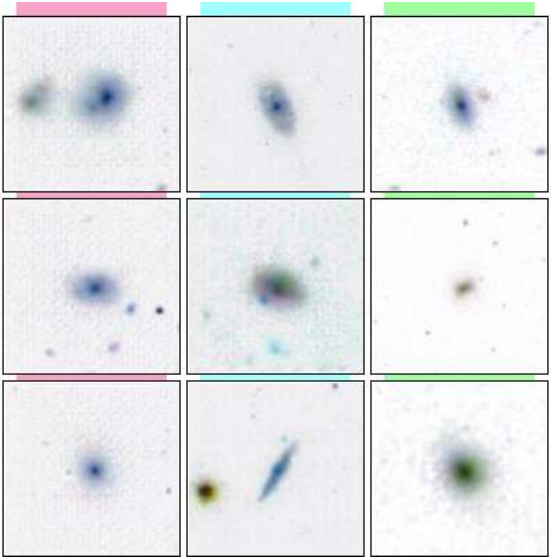}
\centering
\caption{Results for the final architecture to create images with a resolution of \imdim{128}. The first column (red) shows generated images with dropout and a 32-channel disciminator, the second column (blue) shows generated images with dropout removed after 200 epochs, and the third column (green) shows images from the Galaxy Zoo dataset. Images are randomly sampled.}
\label{fig:128Exp}
\end{figure}

As a recommendation for future research, trying an `annealed' dropout in which the probability of dropout starts low and gradually increments to one is an interesting direction, and a generalisation of the training method described above. The results of the final \imdim{128} model are shown in Figure~\ref{fig:128Exp}.

Despite various tests, obtaining the same generation quality with \imdim{128} images as with \imdim{64} images proves to be an obstacle. Although the model produces realistic galaxies, its largest failure is an inability to generate realistic background representations, as shown by the darker colouration of the backgrounds in generated images.

\subsection{StackGAN experiments}
\label{sec:stackganexperiments}

Instead of trying to use a single DCGAN to produce high-quality \imdim{128} images, we use a second GAN, the StackGAN by \citet{stackGan}, to enhance the resolution of realistic images from the \imdim{64} DCGAN to the same resolution realised by \citet{darkGan}. Two datasets are created by scaling down the original \imdim{424} Galaxy Zoo images to a resolution of \imdim{64} and \imdim{128}, respectively, using a nearest-neighbours method. We flip identical pairs of \imdim{64} and \imdim{128} images vertically and horizontally with a probability of $0.5$ for each transformation. The model is then trained for $150$ epochs, with a batch size of $B=64$. While the original StackGAN paper recommends training for $600$ epochs, preliminary experiments show that training for more than $150$ epochs results in mode collapse and a lack of diversity in the generated images. Input images are scaled to be in $[0, 1]$, whereas output images are scaled to be in $[-1, 1]$ as per \citet{superRes1}.

\begin{figure}
\includegraphics[width=\columnwidth]{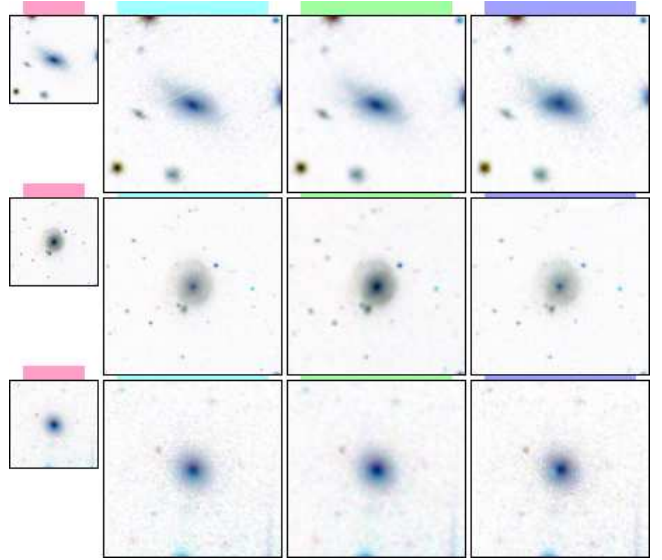}
\centering
\caption{Comparison of resolution-enhanced images with a StackGAN and real images from the Galaxy Zoo dataset. The first column (red) and the second column (blue) show images from the original Galaxy Zoo dataset that are downsampled to a resolution of \imdim{64} and \imdim{128}, respectively. The third column (green) shows images from the Galaxy Zoo dataset with a resolution of \imdim{128} that are upsampled from \imdim{64} versions. The fourth column (purple) shows generated images with a resolution of \imdim{128} conditioned on images of the same resolution from the Galaxy Zoo dataset. Images are randomly sampled.}
\label{fig:stackGanExp}
\end{figure}

The training parameters are specified in Table~\ref{tab:params}, and Figure~\ref{fig:stackGanExp} shows resolution-enhanced results for a sample of generated \imdim{128} images from a random sample of real Galaxy Zoo \imdim{64} images compared to upsampled and downsampled original Galaxy Zoo images.

\begin{figure}
\includegraphics[width=\columnwidth]{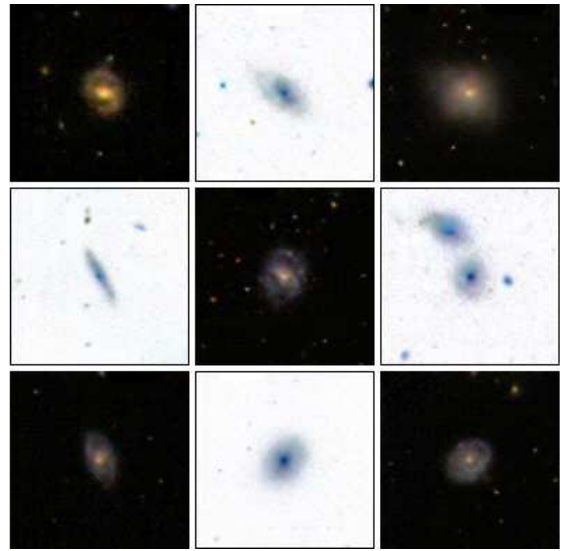}
\centering
\caption{Examples of galaxy images with a resolution of \imdim{128} created with the chained DCGAN/StackGAN model. The images are selected to highlight the model's ability to create features such as spiral arms, as well as a variety of ways in which galaxies present themselves, like edge-on disc galaxies, featureless elliptical galaxies, and multiple galaxies per image.}
\label{fig:typeExamples}
\end{figure}

The results demonstrate the ability of the StackGAN model to generate diverse high-resolution images from lower-resolution synthetic images, and to solve the problem of visible kernel templates described in Section~\ref{sec:generating128}. The architecture presented above, which is comprised of a chained combination of DCGAN and StackGAN, is, therefore, the final model for the generation of \imdim{128} images. Figure~\ref{fig:typeExamples} shows a selection of images generated with the said model.

\section{Evaluation}
\label{sec:eval}

\subsection{Closest-match analysis}
\label{sec:closestmatch}

To show that the trained model does not overfit to the data, random images are sampled from the \imdim{64} DCGAN generator, after which the closest image in the real dataset is found using the $L_1$ distance. For two images $x, x' \in \mathbb{R}^D \times \mathbb{R}^D$, the distance, denoted as $D_{L_1}$, can be expressed as follows:
\begin{eqnarray}
D_{L_1} (x, x') = || x - x' ||_1 = \sum_{i = 1}^D\sum_{j = 1}^D |x^{(i,j)} - x'^{(i,j)}| 
\end{eqnarray}
Before computing the distance, the images are cropped by $50\%$ from the centre to remove the effects of background noise. Through the difference of generated images from the closest-matching real images, we demonstrate that the model learns to create new images from a latent representation instead of memorising and reproducing the dataset.

\begin{figure}
\includegraphics[width=\columnwidth]{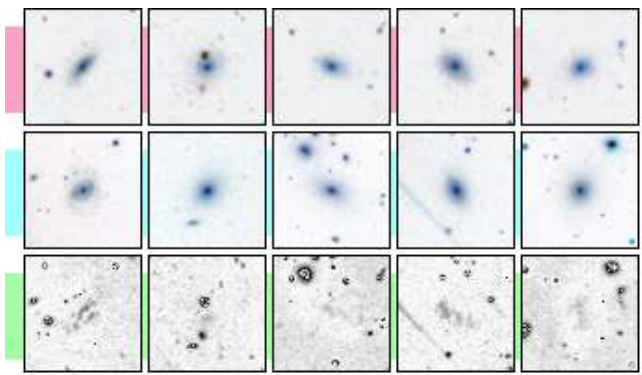}
\centering
\caption{Results for a closest-match analysis to ensure that the generator does not simply memorise the dataset. The first row (red) shows randomly sampled generated galaxy images, the second row (blue) shows the closest-matching real galaxy images from the Galaxy Zoo dataset as measured by the $L_1$ distance, and the third row (green) shows the absolute difference between the two images.}
\label{fig:latentComp}
\end{figure}

By comparing the brightness at the centre of the difference visualisation for pairs of images in the third column of Figure~\ref{fig:latentComp}, it is apparent that the galaxies are not memorised, with discrepancies in colour, shape, and brightness being present between generated images and their closest-matching real counterpart.

\subsection{Property distributions}
\label{sec:propertydistros}

While the generation of visually plausible synthetic galaxy images provides a reasonable proof of concept, the use of generated images in applications within astronomy requires such images to also be physically realistic. Therefore, we assess the quality of generated images by performing statistical tests on both the real and generated data. If the generated statistics closely follow the statistics derived from real data, the generated data is viable for supplementing real datasets in the domain of the measured statistical features.

Through these tests, we explore four properties of the galaxy images: ellipticity, angle of elevation from the horizontal, total flux, and the size measurement of the semi-major axis. Related to this approach, \citet{darkGan} test for two of these properties, ellipticity and size, but with a C-VAE conditioned on the size parameter and in combination with a report that their implementation of a C-GAN produces less consistent results. This introduces the question of whether our trained model can produce consistent evaluation results, which presents an interesting opportunity to compare state-of-the-art results. The ellipticity is defined as:
\begin{eqnarray}
\epsilon = 1 - \frac{s_{\mathrm{major}}}{s_{\mathrm{minor}}}
\end{eqnarray}
Here, $s_{\mathrm{major}}$ and $s_{\mathrm{minor}}$ are the semi-major and semi-minor axis of the ellipse, respectively. We make use of the \textsc{photutils} package, which is part of the \textsc{Astropy} collection of astronomy-related Python packages, to fit an ellipse via isophotes of equal intensity from a predefined elliptical centre \citep{Astropy1, Astropy2}. Specifically, \textsc{photutils} implements methodology initially introduced by \citet{Jedrzejewski1987} to fit measurements around trial ellipses via weighted least-squares. The total flux of the galaxy is then computed as the sum of the pixel values within the outermost ellipse.

While the ellipticity represents a relationship between both axes of an ellipse, the size represents a measure of the semi-major axis. The angle measurement is defined as the angle of elevation relative to the horizontal of the galaxy's semi-major axis in degrees. Due to the limitations of the fitting algorithm in \textsc{Astropy}, all images are upscaled to a resolution of \imdim{512} using bicubic sampling, which allows for more accurate ellipse fits, but does not alter the underlying distribution of the data. The angle of elevation is given in pixels, as the Galaxy Zoo 2 dataset is a $gri$ colour composite of resolution \imdim{424} scaled to $0.02 \cdot \mathtt{petroR90\_r}$ arcseconds per pixel, which means that each image corresponds to a different angular size dependent on the galaxy size. Here, $\mathtt{petroR90\_r}$ denotes the radius containing $90\%$ of the $r$-band Petrosian aperture flux \citep{galaxyZoo2}. For the same reason, the total fluxes are treated as relative fluxes due to the lack of a consistent conversion from flux per pixel to flux per angle.

\begin{figure}
\includegraphics[width=\columnwidth]{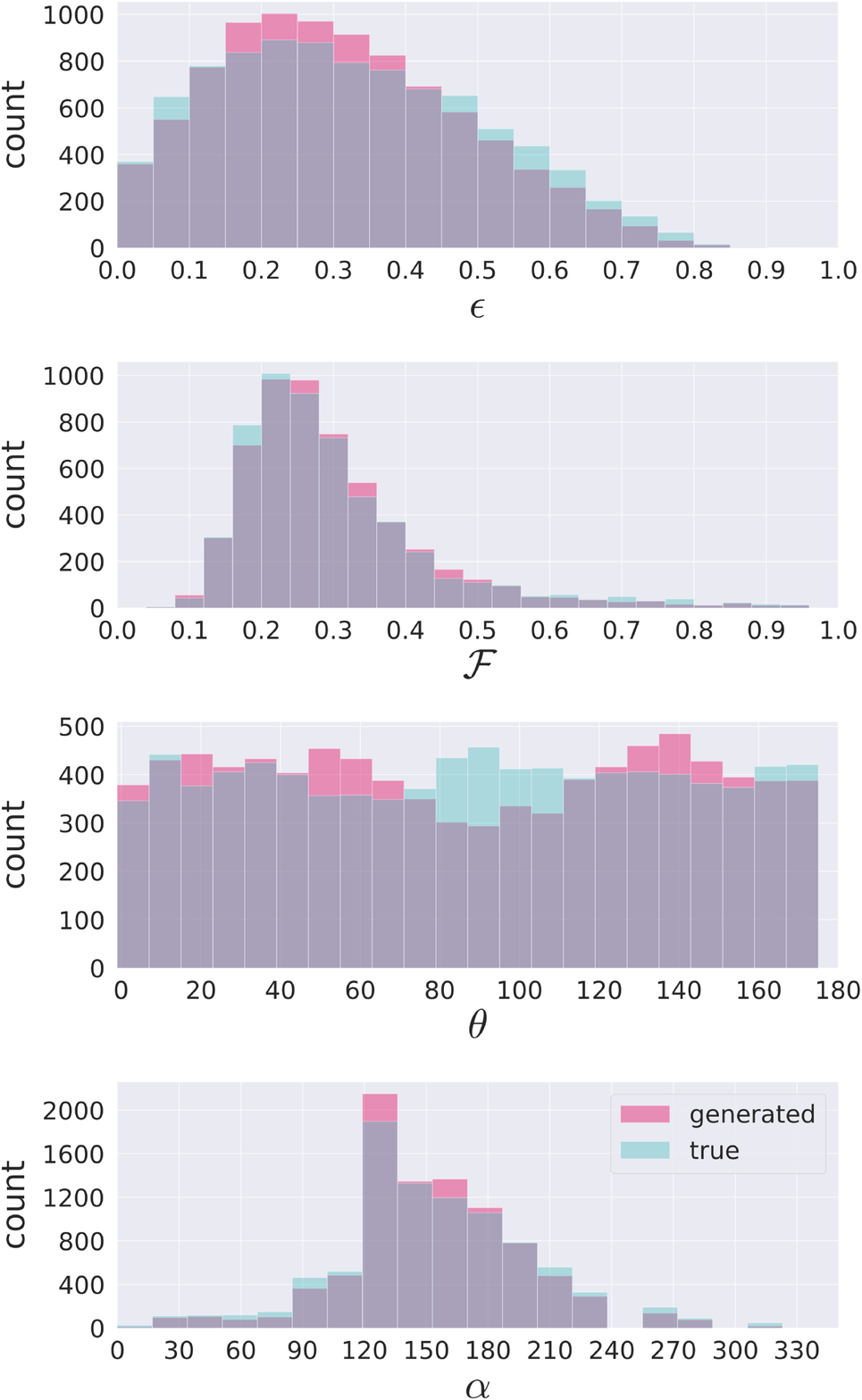}
\centering
\caption{Histograms of the evaluation metrics for synthetic galaxy images with a resolution of \imdim{64} created with a DCGAN. The distributions for generated images and real Galaxy Zoo dataset samples, both upsampled from \imdim{64} to \imdim{512}, are coloured in red and blue, respectively. The first row shows the distributions for the ellipticities ($\epsilon$) of the images, the second row shows the distributions for angles of elevation from the horizontal in degrees ($\theta$), the third row shows the distributions for relative fluxes ($\mathcal{F}$), and the fourth row shows the distributions for the size measured by the semi-major axis in pixels ($a$). All plots are created with random samples of 9,000 images from both the generated images and the Galaxy Zoo dataset.}
\label{fig:dis64Plot}
\end{figure}

\begin{figure}
\includegraphics[width=\columnwidth]{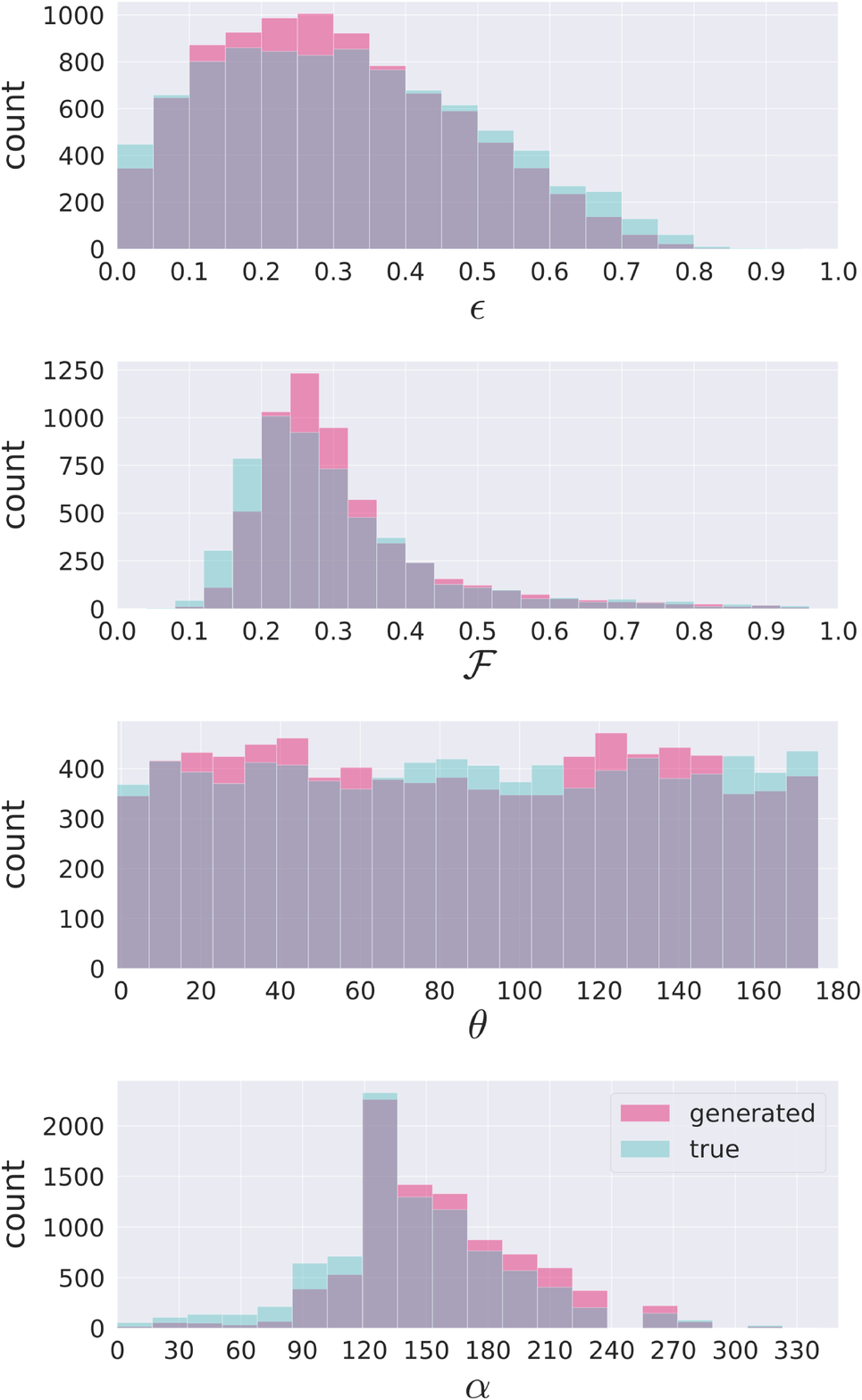}
\centering
\caption{Histograms of the evaluation metrics for synthetic galaxy images with a resolution of \imdim{128} created with a two-stage pipeline using a DCGAN for \imdim{64} images and then upscaling the images to a higher resolution of \imdim{128} with a StackGAN. The distributions for generated images and real Galaxy Zoo dataset samples, both upsampled from \imdim{128} to \imdim{512}, are coloured in red and blue, respectively. The first row shows the distributions for the ellipticities ($\epsilon$) of the images, the second row shows the distributions for angles of elevation from the horizontal in degrees ($\theta$), the third row shows the distributions for relative fluxes ($\mathcal{F}$), and the fourth row shows the distributions for the size measured by the semi-major axis in pixels ($a$). All plots are created with random samples of 9,000 images from both the generated images and the Galaxy Zoo dataset.}
\label{fig:dis128Plot}
\end{figure}

The four statistics for each galaxy image are measured for a sample of $9,000$ galaxy images from both the true data set and the generated data. How well the generated data incorporates key galaxy features can be measured by comparing the distribution of the statistics over the real and generated data. Evaluations are performed on the best \imdim{64} image generative model and the best \imdim{128} image generative model. Figure~\ref{fig:dis64Plot} shows a comparison of generated and true distributions for each of the four statistics, with generated images for a resolution of \imdim{64} from our DCGAN being evaluated. Similarly, Figure~\ref{fig:dis128Plot} depicts the same comparison plots for a resolution of \imdim{128}, with the upscaled images obtained from the two-stage generation process using our chained combination of DCGAN and StackGAN.

The ellipse fitting via \textsc{photutils} fails to fit suitable ellipses for some of the generated images, and the \textsc{Astropy} library states that: ``A well defined negative radial intensity gradient across the region being fitted is paramount for the achievement of stable solutions''\footnote{\href{http://photutils.readthedocs.io/en/stable/api/photutils.isophote.Ellipse.html}{\nolinkurl{http://photutils.readthedocs.io/en/stable/api/} \newline \nolinkurl{photutils.isophote.Ellipse.html}}}. Increasing the scale of the images helps reduce the percentage of ellipse-fitting failures, and we observe that the similarity of the distributions is inversely proportional to the percentage of failed fits. For the final results of the \imdim{64} resolution distributions, approximately $5\%$ of ellipse fits failed, which is likely to be the cause of the `dip' of the generated distribution of Figure~\ref{fig:dis64Plot} in the angle evaluation in the distribution comparisons, shifting mass to the extremes of the distribution.
While this primarily concerns the plotting for evaluation, non-uniform orientations in generated datasets can be easily fixed by randomly rotating images. In contrast, failures to fit did not occur for the \imdim{128} resolution images generated by the StackGAN, thus correcting a lack of a defined negative radial intensity gradient in a fraction of the generated images and, as a result, the dip in the angle distribution.

As can be seen in Figure~\ref{fig:dis64Plot} and Figure~~\ref{fig:dis128Plot}, apart from slight incongruities in the angle plot for the \imdim{64} DCGAN and the flux plot for the final \imdim{128} chained DCGAN/StackGAN model, the distributions of the investigated properties for the generated data closely follow the distributions for the real data, demonstrating that the model has learned an effective latent representation of galaxy features. Given that our model is not conditioned on any galaxy features, this confirms the general viability of our approach for the types of applications discussed in Section~\ref{sec:intro}.

\section{Discussion}
\label{sec:discussion}

One interesting finding of the experiments and their accompanying results is that a comparatively simple model produces the most realistic images as well as the best evaluation results. Despite experimenting with the use of dropout and label smoothing, the original DCGAN architecture outperformed all other models on the dataset with resolution \imdim{64}, hence supporting the case for model simplicity. This outcome is in line with recent debates on whether neural networks memorise the data they are trained on, with results supporting both sides being presented \citep{nnMem1, nnMem2}. \citet{allConv} show that techniques such as simplifying a model's architecture can improve test results in such a way that smaller models are competitive with state-of-the-art models of higher complexity, with similar findings being reported about adjustments in the initialisation \citep{xavierInit, goodInit}. As described before, the dataset with a resolution of \imdim{128} requires a customised architecture to reach the model's best performance, but those changes do not represent a significant deviation from the DCGAN model that generates the initial galaxy images that are then resolution-enhanced.

Our statistical evaluation of physical properties in the generated images also shows that our models are able to learn realistic latent representations of data. This requires both the discriminator and the generator to extrapolate these underlying features defining the property distributions of galaxies in our universe, only through backpropagation. We find that models tasked with directly generating larger images with a resolution of \imdim{128} struggle primarily with filling the background around the galaxies. One solution for this case is to crop the galaxy images by $50\%$ as done by \citet{darkGan}, which removes the need for the model to learn complex background-filling techniques. Building from this pre-processing, a potential enhancement is to encourage the discriminator to focus on different sub-regions of the generated and real images by cropping random sub-sections of the input data. We do, however, find that an effective approach is to train a second generator to increase the resolution of the generated \imdim{64} images, avoiding the need for pre-processing techniques. One might also consider whether the statistical evaluation of the physical properties in the generated images could be used to enhance the quality of the generator, similar to the addition of the pixel-loss term in Section \ref{stackgan}, by defining a loss metric such as the Kullback-Leibler divergence to minimise the difference between the real and generated evaluation distributions. Although this technique would certainly improve the quality of the generated distributions relative to the evaluation criteria, a bias would inherently be introduced into the model. Because the evaluation criteria would no longer be independent of the model's objective, evaluating the model according to the original evaluation metrics would be problematic. This is a case of Goodhart's Law: ``When a measure becomes a target, it ceases to be a good measure'' \citet{GoodhartsLaw}. To overcome this issue, further evaluation metrics would have to be defined such that we can determine the quality of our model without bias.

As discussed in Section~\ref{sec:propertydistros}, the distributions for angles of elevation from the horizontal deviate slightly from the real dataset for the \imdim{64} DCGAN model, which can be resolved by randomly flipping generated images along the axes. In contrast, the evaluation for the \imdim{128} chained DCGAN/StackGAN model is shown to closely follow the distributions for the real dataset, the exception being the flux evaluation due to generated galaxies being slightly brighter on average. An interesting characteristic of the StackGAN model is that it does not just enhance the resolution quality, but is also able to sometimes correct defects in the original galaxy images, as can be seen in the first row of Figure~\ref{fig:stackGanExp}. This model behaviour can be ascribed to the dual-objective function seen in equation~\ref{eq:dual_obj}. Specifically, the generative loss term corrects real galaxy images that have artifacts which do not match the galaxies' distribution.

Generative models that are able to create physically realistic galaxy images have many practical uses. In this work, we use Galaxy Zoo 2 images, which are $gri$ compound images with the bands fed into RGB channels. Our described architectures can, however, easily be used for different bands and numbers of channels. The Galaxy Zoo datasets are hand-classified via community crowd sourcing, and thus features such as shape, merging disturbance, and irregularities are determined through a hand-built decision tree. Using this manual classification approach on the generated data would provide a more rigorous evaluation metric for the model, for example through measurements of the distributions for different classes, as well as an assessment of the ease of classification, both in comparison to the real datasets. 

Another obvious use of models such as the ones presented in this paper is the creation of large datasets of high-quality galaxy images that are representative of a specific survey. Shape measurement algorithms used to detect weak lensing signals are an important part of research targeting dark matter, for example in the context of upcoming surveys like LSST and Euclid \citep{lsst, euclid}. The process of calibrating for measurement biases relies on image simulations with a known ground truth, which requires high-quality images as an input to the simulation. For this reason, the distributions of generated data used in place of expensive observations have to closely follow the distributions of real data for properties such as ellipticity. Scaling the generated images to a larger resolution would make the results more applicable to a wider range of astronomical applications. We propose that higher-resolution images could be obtained by using a larger model. There are two possible approaches to this: One approach is to continue the chaining of GANs by scaling up the \imdim{128} images to the target resolution with another resolution-enhancing GAN. This might exacerbate artifacts from the smaller models, so training these additional GANs on real as well as fake images could be necessary. A second approach is to keep the current StackGAN model unchanged and, alternatively, train the model to generate \imdim{128} cropped segments of the larger target image, conditioned on the coordinate of the image, similar to the approach in \citet{superRes1}. To generate, for example, a \imdim{512} image, the model would first generate sixteen \imdim{128} image segments and then combine these segments to create a final image. Both of these approaches do, however, require copious computational resources.

With regard to weak lensing, the primary proposed application of our method is the calibration of shape measurement algorithms, either with generated galaxy images used as a ground truth to which biases can be applied, or to check whether the same kind of biases can be detected in both real and generated images. As weak lensing measurements are taken in aggregate, stacked information from ensembles of images can additionally be used for calculating evaluation statistics to compare the generated and real images in terms of features that are too weak to be inferred from individual samples. Specifically, we propose a GREAT3-style comparison, as described in \citet{Mandelbaum2014}, of shape statistics measured on the training sample to those measured on generated images. Multiplicative and additive biases that are retrieved for both generated and real samples can then be compared to further test the suitability for shape measurement calibration.

Finally, a natural extension of the presented research is to test the effect that augmenting datasets from efforts such as Galaxy Zoo or EFIGI has on deep object-segmentation or galaxy classification models. Showing that generated galaxy images improve the generalisation and test accuracy of these models motivates further research into deep learning models for astronomy.

\section{Conclusion}
\label{sec:conclusion}

In this paper, we show how generative modelling with GAN architectures can be used for the augmentation of smaller datasets of galaxy images. Specifically, the original DCGAN architecture proves sufficient for the generative model to create physically realistic images that closely follow the property distributions of real galaxy images when faced with statistical evaluations.

In addition, we explore the applicability and limits of common ways to optimise such models, and show that the StackGAN architecture can be used as a second-stage model in a chained DCGAN/StackGAN approach to generate synthetic galaxies with higher resolutions, circumventing the difficulties that DCGAN models experience with such resolutions. While GANs have quickly spread to a variety of application areas since their introduction in 2014, our work also adds to the evidence that chaining different GAN models is a workable approach.

By demonstrating that distributions of generated galaxies closely follow the real data distributions for a variety of physical properties, we propose that the generated galaxy images can be used to augment real galaxy datasets and enlarge the number of samples from surveys. The range of evaluation metrics used in this paper show the viability of synthetic galaxies generated in this way for learning tasks such as galaxy classification and segmentation, deblending, and the calibration of shape measurement algorithms used to investigate dark energy through weak gravitational lensing. With the presented capability to provide a data source for deep learning models that require a large number of training samples, our work demonstrates the potential of GAN architectures as a valuable tool for modern-day astronomy.

\section*{Acknowledgements} 

We would like to express our gratitude to the Galaxy Zoo team for creating a high-quality dataset of galaxy images well-suited for machine learning applications, the anisotropic irregularities in the early universe for making them possible in the first place, and the University of Edinburgh's School of Informatics for supplying the necessary GPU power. We also wish to thank Romeel Dav\'e and Joe Zuntz for helpful conversations about evaluation metrics and discussions about explanatory requirements, and Nathan Bourne for coordinating the summer research projects at the University of Edinburgh's Institute for Astronomy during the 2017/2018 period, which laid the foundation for this paper.

\bibliographystyle{mnras}
\bibliography{ref}

\appendix

\section{}
\label{app:appA}

Both the generator and discriminator in the \imdim{64} DCGAN are composed of five layers. The generator increases the image dimensionality from $D_0=1$ to $D_1=4$, then from $D_1=4$ to $D_2=8$, then from $D_2=8$ to $D_3=16$, then from $D_3=16$ to $D_4=32$, and finally from $D_4=32$ to $D_5=64$. Each layer is composed of a deconvolutional component of dimensions $C \times 4 \times 4$, where $C$ denotes the channel size. The number of channels starts at $C=1024$ and halves each layer.

\vspace{10pt}

\noindent\fbox{%
\parbox{0.965\columnwidth}{%
\noindent \textbf{64$\times$64 DCGAN generator}
\vspace{5pt}
\begin{small}
\begin{enumerate}[(1)]
\setlength\itemsep{0.1em}
\item 100-dimensional multivariate Gaussian input
\item $L = 5$ deconvolution layers:
\begin{footnotesize}
\begin{description}
\item Channel size: $2^{x_i}$ with $x \coloneqq \{10, 9, 8, 7,  \frac{\ln 3}{\ln 2}\}$
\item Padding: $\{0,1,1,1,1\}$
\item Stride: $\{1,2,2,2,2\}$
\end{description}
\end{footnotesize}
\item Batch size: 32
\item Kernel size: $4\times4$
\item Batch normalisation after layers $l \notin \{1, L\}$
\item ReLU activation function in layers $l \neq L$
\item Tanh activation function in layer $l = L$
\end{enumerate}
\end{small}
}%
}

\vspace{10pt}

The deconvolution is followed by a batch normalisation component, described in  Section~\ref{sec:batchnorm}, and concludes with a rectified linear unit (ReLU) activation function defined as:
\begin{eqnarray}
\label{eq:relu}
\varphi(x) = \text{max}(0, x) 
\end{eqnarray}
Here, the input $x$ is defined as a kernel convolved over a $C \times 4 \times 4$ patch of the input into the layer:
\begin{eqnarray}
\label{eq:inputdef}
x_{i,j,k} = \sum_{l=0}^4 \sum_{m=0}^4 \sum_{p=0}^C w_{l,m,p} a_{i+l,j+m,k+p}
\end{eqnarray}
Instead of a ReLU activation, the output layer has a tanh activation function defined as:
\begin{eqnarray}
\label{eq:tanh}
\varphi(x) = \frac{1-e^{-2x}}{1+e^{-2x}}
\end{eqnarray}
The input and output layers do not have batch normalisation. By setting the padding to $p=1$, the stride to $s=2$, and the kernel size to $k=4$, the doubling of the image dimensions is enforced by the deconvolution equation:
\begin{eqnarray}
\label{eq:deconv}
D_{l+1} = (D_{l}-1)s + k - 2p
\end{eqnarray}
This equation shows that the dimensionality of the image in layer $D_{l+1}$ is a function of the dimensionality of the image in layer $D_l$. The exception is the input layer, which is padded by $p=0$ and has a stride of $s=1$ due to the input image dimension being $D_0=1$, hence resulting in an output size of $D_1=4$. 

The disciminator is almost identical to the generator, but inverts the image process by scaling the image from $D_0=64$ to $D_5=1$ by halving the resolution in each layer. 

\vspace{10pt}

\noindent\fbox{%
\parbox{0.965\columnwidth}{%
\noindent \textbf{64$\times$64 DCGAN discriminator}
\begin{small}
\vspace{5pt}
\begin{enumerate}[(1)]
\setlength\itemsep{0.1em}
\item $L = 5$ convolution layers:
\begin{footnotesize}
\begin{description}
\item Channel size: $2^{x_i}$ with $x \coloneqq \{7, 8, 9, 10, 0\}$
\item Padding: $\{1,1,1,1,0\}$
\item Stride: $\{2,2,2,2,1\}$
\end{description}
\end{footnotesize}
\item Batch size: 32
\item Kernel size: $4\times4$
\item Batch normalisation after layers $l \notin \{1, L\}$
\item LeakyReLU activation function in layers $l \neq L$
\item Sigmoid activation function in layer $l = L$
\end{enumerate}
\end{small}
}%
}

\vspace{10pt}

Instead of rectified linear units, leaky rectified linear units (LReLU) are used after the batchnorm, defined as:
\begin{eqnarray}
\label{eq:lrelu}
\varphi(x)=\mathbb{I}\{x < 0\} \alpha x + \mathbb{I}\{x \geq 0\}x
\end{eqnarray}
Here, $\mathbb{I}\{\lambda\}=1$ is the case if statement $\lambda$ is true, and $\mathbb{I}\{\lambda\}=0$ otherwise, while $\alpha$ is the leaking rate which is set to $\alpha=0.2$, and $x$ is defined as in Equation~\ref{eq:inputdef}. The output layer has a sigmoid activation function defined as:
\begin{eqnarray}
\label{eq:sigmoid}
\varphi(x)=(1+e^{-x})^{-1}
\end{eqnarray}
The equation for determining the dimensionality of the output of a layer's convolution is the inverse of the deconvolution equation:
\begin{eqnarray}
\label{eq:conv}
D_{l+1} = \frac{(D_{l} - k + 2p)}{s} + 1
\end{eqnarray}
Thus, the last layer has a padding of $p=0$ and a stride of $s=1$, like the first layer in the generator, so that the final layer has output dimensionality $D_5=1$. The number of channels of the convolutional layers in the discriminator starts at $C=128$ and doubles with every layer.

\section{}
\label{app:appB}

The StackGAN Stage-II generator takes as its input a \imdim{64} image. This is first downsampled to \imdim{32} by a convolutional layer with $C=128$ convolution channels of size $3 \times 4 \times 4$, a padding of $p=1$, and a stride of $s=2$, after which a series of six residual blocks follows. Each residual block consists of two convolutional layers with stride $s=1$, kernel size $k=3$, padding $p=1$, and channel size $C=128$, so that the image resolution remains \imdim{32} through these layers according to Equation~\ref{eq:conv}.
Batch normalisation, as described in Section~\ref{sec:batchnorm}, and a rectified linear unit activation as seen in Equation~\ref{eq:relu}, are placed after each convolutional layer.

\vspace{10pt}

\noindent\fbox{%
\parbox{0.965\columnwidth}{%
\noindent \textbf{Stage-II StackGAN generator}
\vspace{5pt}
\begin{small}
\begin{enumerate}[(1)]
\setlength\itemsep{0.1em}
\item \imdim{64} image input
\item $L = 16$ convolution layers:
\begin{footnotesize}
\begin{description}
\item Kernel size: \imdim{4} for layers $l = 1$, \imdim{3} otherwise
\item Channel size: $3 \rightarrow 128$ for layer $l = 1$\\
\phantom{\hspace{7.05em}}$128 \rightarrow 128$ for layers $l \notin \{1,L - 1,L\}$\\
\phantom{\hspace{7.05em}}$128 \rightarrow 512$ for layer $l = L - 1$\\
\phantom{\hspace{7.15em}}$512 \rightarrow 3$ for layer $l = L$
\item Padding: $1$ for all layers $l \in \{1,\ldots,L\}$
\item Stride: $1$ for layers $l \neq 1$, $2$ otherwise
\end{description}
\end{footnotesize}
\item Batch normalisation after layers $l \notin \{1, L\}$
\item ReLU activation function in layers $l \neq L$
\item Tanh activation function in layer $l = L$
\item Upsampling ($\times 2$) before layers $l \in \{L - 1, L\}$
\item Residual connections:
\begin{footnotesize}
\begin{description}
\item Add layer $l = 1$ output to layer $l = L - 2$ output
\item Add layer $l \in \{2n \ | \ n \in \mathbb{N}_{<7} \setminus \{0\}\}$ input\\
\phantom{\hspace{1.75em}} to layer $l \in \{2n + 1 \ | \ n \in \mathbb{N}_{<7} \setminus \{0\}\}$ output
\end{description}
\end{footnotesize}
\end{enumerate}
\end{small}
}%
}

\vspace{10pt}

Importantly, the residual block adds its input to the batch-normalised output of the final convolutional layer via a simple sum operation before passing both through the activation function: 
\begin{eqnarray}
\label{eq:resblock}
a_{l+1} = \varphi(b_{l} + a_{l-1})
\end{eqnarray}
Here, $a_{l-1}$ is the input to the residual block, $a_{l+1}$ is the output of the residual block, and $b_{l}$ is the result of the convolution and batchnormalisation stages in the residual block. After the six residual blocks, another residual convolutional layer is added, which incorporates skip connections from the first layer of the generator as described by:
\begin{eqnarray}
\label{eq:rescon}
a_{8} = \varphi(b_{8} + a_{1})
\end{eqnarray}
In the above equation, $a_{1}$ is the output from the first downsampling layer of the network, $a_{8}$ is the output of the residual connections in the eighth layer, and $b_{8}$ is the result of the convolution and batchnorm stages in the eighth layer, prior to the activation. The model then has two upsampling layers, which double the image resolution using nearest-neighbours pixel sampling and where each pixel $x_{i,j}$ in the $D_{l+1}=2D_{l}$ image is assigned the value from its nearest neighbour in the original $D_{l}$ image with pixels $x'_{i,j}$ according to $x_{i,j}=x'_{i',j'}$ where:
\begin{eqnarray}
\label{eq:nnup}
i',j' = \argmin_{k,m \in D_{l}} \{|\beta k-i|+|\beta m-j|\}
\end{eqnarray}
Here, $\beta=2$ is the upscaling factor. A stride $s=1$, kernel size $k=3$, and padding $p=1$ convolutional layer are then placed after the upsampling layer to enable the network to make image adjustments without changing the image resolution. The first upsampling layer has a convolution channel size of $C=128$, a batch normalisation, and rectified linear unit activation after it; the second upsampling layer has a convolutional channel size of $C=512$ followed by a tanh activation function as defined in Equation~\ref{eq:tanh}.

\section{}
\label{app:appC}

A full description of the \imdim{64} DCGAN generator architecture is provided in Appendix~\ref{app:appA}. The \imdim{128} DCGAN architecture extends the \imdim{64} DCGAN architecture by adding an additional deconvolutional layer in the generator before the tanh activation, so that the images are enhanced from \imdim{64} to \imdim{128}.

\vspace{10pt}

\noindent\fbox{%
\parbox{0.965\columnwidth}{%
\noindent \textbf{128$\times$128 DCGAN generator}
\vspace{5pt}
\begin{small}
\begin{enumerate}[(1)]
\setlength\itemsep{0.1em}
\item 100-dimensional multivariate Gaussian input
\item $L = 6$ deconvolution layers:
\begin{footnotesize}
\begin{description}
\item Channel size: $2^{x_i}$ with $x \coloneqq \{11, 10, 9, 8, 7,  \frac{\ln 3}{\ln 2}\}$
\item Padding: $\{0,1,1,1,1,3\}$
\item Stride: $\{1,2,2,2,2,2\}$
\end{description}
\end{footnotesize}
\item Batch size: 32
\item Kernel size: $4\times4$
\item Batch normalisation after layers $l \notin \{1, L\}$
\item ReLU activation function in layers $l \neq L$
\item Tanh activation function in layer $l = L$
\end{enumerate}
\end{small}
}%
}

\vspace{10pt}

In Appendix~\ref{app:appA}, a full description of the \imdim{64} DCGAN discriminator architecture is given. The \imdim{128} DCGAN architecture extends the \imdim{64} DCGAN architecture by adding an additional convolutional layer in the discriminator before the first input layer, so that the images are down-sampled from \imdim{128} to \imdim{64}. Additionally, a dropout layer, as described in Section~\ref{sec:dropout}, is added before the final convolutional layer for the first $200$ epochs.

\vspace{10pt}

\noindent\fbox{%
\parbox{0.965\columnwidth}{%
\noindent \textbf{128$\times$128 DCGAN discriminator}
\begin{small}
\vspace{5pt}
\begin{enumerate}[(1)]
\setlength\itemsep{0.1em}
\item $L = 6$ deconvolution layers:
\begin{footnotesize}
\begin{description}
\item Channel size: $2^{x_i}$ with $x \coloneqq \{5, 6, 7, 8, 9, 0\}$
\item Padding: $\{3,1,1,1,1,0\}$
\item Stride: $\{2,2,2,2,2,1\}$
\end{description}
\end{footnotesize}
\item Batch size: 32
\item Kernel size: $4\times4$
\item Batch normalisation after layers $l \notin \{1, L\}$
\item LeakyReLU activation function in layers $l \neq L$
\item Sigmoid activation function in layer $l = L$
\item Dropout in layer $l = L - 1$ for epochs $e \in \{1, 2,\ldots, 200\}$
\end{enumerate}
\end{small}
}%
}

\bsp

\label{lastpage}
\end{document}